\documentclass[twocolumn, showpacs, preprintnumbers, amsmath, amssymb]{revtex4}
\usepackage{dcolumn}
\usepackage{graphicx}
\usepackage{bm}
\begin{document}
\title{Fracture Roughness and Correlation Length in the Central Force Model}
\author{Jan {\O}ystein Haavig Bakke}
\email{Jan.Bakke@ntnu.no}
\author{Thomas Ramstad}
\email{Thomas.Ramstad@ntnu.no}
\author{Alex Hansen}
\email{Alex.Hansen@ntnu.no} 
\affiliation{Department of Physics, Norwegian University of Science and
Technology, N--7491 Trondheim, Norway}
\date{\today}
\begin{abstract}
We measure the roughness exponent and the correlation length exponent 
of a stress-weighted percolation process in the central force model in 2D.  
The roughness exponent is found to be $\zeta = 0.75 \pm 0.03$ and the 
correlation length exponent is found to be $\nu = 1.7 \pm 0.3$.  This 
result supports a conjecture that the fracture roughness for large scales 
is controlled by a stress weighted percolation process, and the fracture 
roughness can by calculated from the correlation length exponent by 
$\zeta = 2\nu/(1+2\nu)$.  We also compare global and local 
measurements of the fracture roughness and do not find sign of anomalous 
scaling in the central force model.  
\end{abstract}
\maketitle
In the early eighties, it was observed that brittle fracture surfaces 
show self affinity.  That is, such a surface will be statistically invariant
if the in-plane length scales are changed by a factor $\lambda$, and the
out-of plane length scale is changed by a factor $\lambda^\zeta$, where
$\zeta$ is the roughness or Hurst exponent \cite{mpp84-bs85}.  In the
early nineties, it was proposed that the roughness exponent is universal
\cite{blp90}.  This initiated a large effort to further investigate this
phenomenon \cite{mhhr92-sgr93-cw93-khw93-emhr94-sss95}. An 
early review may be found in Ref.\ \cite{b97}.

For large scales the universal value of the roughness exponent is believed to 
be about 0.80 for brittle fractures.  There is also evidence that there may be
a different value, 0.5,
for small scales, with a well-defined crossover length scale separating
these two regimes \cite{dhbc96-dnbc97}.

Even though the evidence for one or more universal roughness exponents is
mounting, the mechanisms that may be responsible are still not known.  There
have been some proposals for mechanisms, see \cite{bblp93,bb94,bhlp02,hs03}.
It is the aim of this Letter to test the basic mechanism proposed by Hansen
and Schmittbuhl \cite{hs03}, connecting the roughness exponent to a 
correlation length exponent associated with the underlying fracture process. 
The basic idea is that the breakdown process when the material is very 
disordered is essentially a correlated percolation-like process --- percolation
like in the sense that the machinery for describing it is the same as in 
ordinary percolation.  For smaller disorders, the breakdown process localizes,
and by combining the localization length with the percolation-like
description, a relation may be set up between the roughness exponent $\zeta$
and the correlation length exponent $\nu$ of the correlated percolation
process,  
\begin{equation}
  \zeta = \frac{2\nu}{1+2\nu}\;.
  \label{eq:sh}
\end{equation}
This relation has been tested on the two and three-dimensional fuse model
\cite{bbrshs03,rbbsh04}.  The fuse model consists of a regular network of
electrical fuses, each having a burn-out threshold drawn from some spatially
uncorrelated statistical distribution \cite{arh85,hr90}.  The model has
been used quite extensively to study fracture roughness 
\cite{hhr91,bh98,rsad98,sra00,hs03}.  In this Letter, we test Eq.\ 
(\ref{eq:sh}) in the central-force breakdown model, first studied in
Ref.\ \cite{hrh89}.  This is a model that is much closer to the fracture
problem in that its response is elastic --- in the continuum limit it
maps onto ordinary Lam{\'e} elasticity.  It consists of a regular lattice
whose bonds are elastic springs free to rotate around the nodes they are
connected to.  The force on a spring between nodes   
$\bm{r}_i$ and $\bm{r}_j$ given as \cite{fs84}
\begin{equation}
  f_{ij} = \sigma_{ij}\left(\bm{r}_j-\bm{r}_i)\right)\cdot \bm{n}_{ij}\cdot 
\bm{n}_{ij}\;,
  \label{eq:cfforce}
\end{equation}
where $\sigma_{ij}$ is the spring constant for the spring and $\bm{n}$ is the 
axis vector of the spring.  Another elastic model that has been studied is
the beam model \cite{hhr89,shh01}.

Measuring both $\zeta$ and $\nu$ independently in the two-dimensional 
central-force breakdown model, we will be able to test Eq.\ (\ref{eq:sh}) for 
this case.  For pure central-force rigidity percolation in two dimensions, 
Moukarzel and Duxbury \cite{md99} found $\nu = 1.16$ using the pebble game 
algorithm.  As is the case for the fuse model \cite{rbbsh04,knsz05}, 
there is no reason whatsoever that the $\nu$ of the breakdown process 
should be equal to the percolation correlation exponent.  In the fuse model,
one finds $\nu=1.56$ in two dimensions to be compared to $\nu=4/3$ in
two-dimensional percolation, and $\nu=0.83\pm0.04$ in the three-dimensional
fuse model, to be compared to $\nu=0.88$ for three-dimensional percolation.  
As we shall see, we find $\nu=1.7\pm0.3$ for the two-dimensional central-force
breakdown model.

We simulated fractures in 2D trigonal central force lattice with periodic 
boundaries in the $x$-direction and applied tension in the $y$-direction. 
Each spring in the lattice was assigned a threshold $t_i$ at random from a 
power law distribution $\pi(t) \propto t^{-1+\beta}$ which gives 
$t_i = r_i^D$ where $r_i$ is a random number between 0 and 1 and 
$D =\beta^{-1}$.  Values of $D$ close to zero gives narrow disorders. 
The disorder becomes broader when $D$ increases.  We used $D = 0.7$ for the 
narrow disorder and $D = 20$ for the broad disorder. For each realisation, 
a strain of unity is applied in the vertical direction to the boundary 
lattice points and the central force equations are solved iteratively with 
the conjugate gradient method \cite{bh88}. 
The lattice is periodic in the horizontal direction. The spring with 
$\max_{i}(f_i/t_i)$ is identified and the strain is increased (decreased) 
until $f_i/t_i = 1$ and the spring constant for this spring is set to zero. 
The central force equations are solved again with this new configuration 
and springs are removed until the elasticity module of the system is zero.

\begin{figure}
\includegraphics[width=7cm]{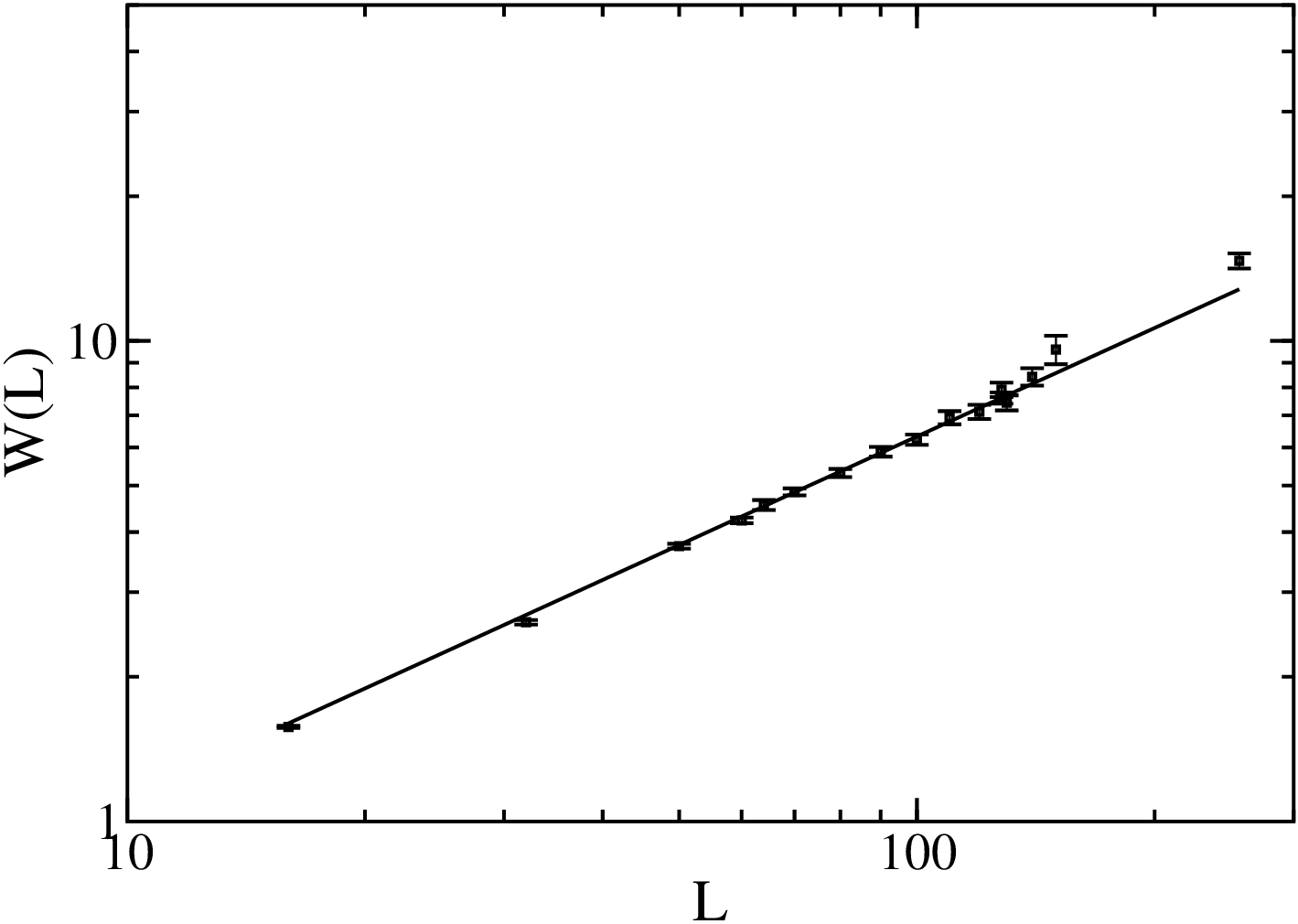}
  \caption{\label{fig:rmswidth}The global roughness exponent is found by 
considering the scaling of the average fracture width of a fracture profile 
$y(x)$, using Eq.\ (\ref{eq:wlscale}),
where the averages are done over the whole system. 
The solid line is $\zeta = 0.75$. We used from 5499 samples for 
$L = 16$ to 237 samples for $L = 256$.}
\end{figure}

When a sample is fractured, elasticity module equal zero, 
the fracture surface is created in the dual lattice 
which gives $2\times L_x$ measuring points for the fracture 
surface for the triangular lattice. Before the 
fracture width is calculated, overhangs are removed using a solid-on-solid 
approximation.  

In order to address the question of whether there is a possibility for 
{\it anomalous scaling\/} of the roughness \cite{lrc97,ls98}, 
we distinguish between
measuring the roughness exponents using global methods --- that is, when the
roughness is measured along the entire length of the system, and the system
size is changed, and local methods, where the size of a window is changed
whilst the system size is kept fixed. When global and local methods give
different roughness exponents, the system is anomalously rough.

We calculate the global fracture roughness exponent by finding how the 
fracture width scales with system size 
\begin{equation}
  W(L) = (\langle y(x)^2\rangle_L - 
\langle y(x))\rangle_L^2)^{1/2} \propto L^{\zeta}\;,
  \label{eq:wlscale}
\end{equation}
and the local fracture roughness exponent with the scaling of the local width
with window size
\begin{equation}
  w(l) = (\langle y(x)^2\rangle _l - \langle y(x)\rangle^2_l)^{1/2} 
\propto L^{\zeta_{loc}}\;,
  \label{eq:lwlscale}
\end{equation}
and the average wavelet coefficients (AWC) method \cite{shn98}
\begin{equation}
  W[y](a) \propto a^{\zeta_{loc} + 1/2}\;.
  \label{eq:wascale}
\end{equation}

The simulations for measuring the fracture roughness was done with a narrow 
disorder of $D = 0.70$. For the global exponent we measured 
$\zeta = 0.75 \pm ± 0.03$.

\begin{figure}[tbp!]
  \includegraphics[width=7cm]{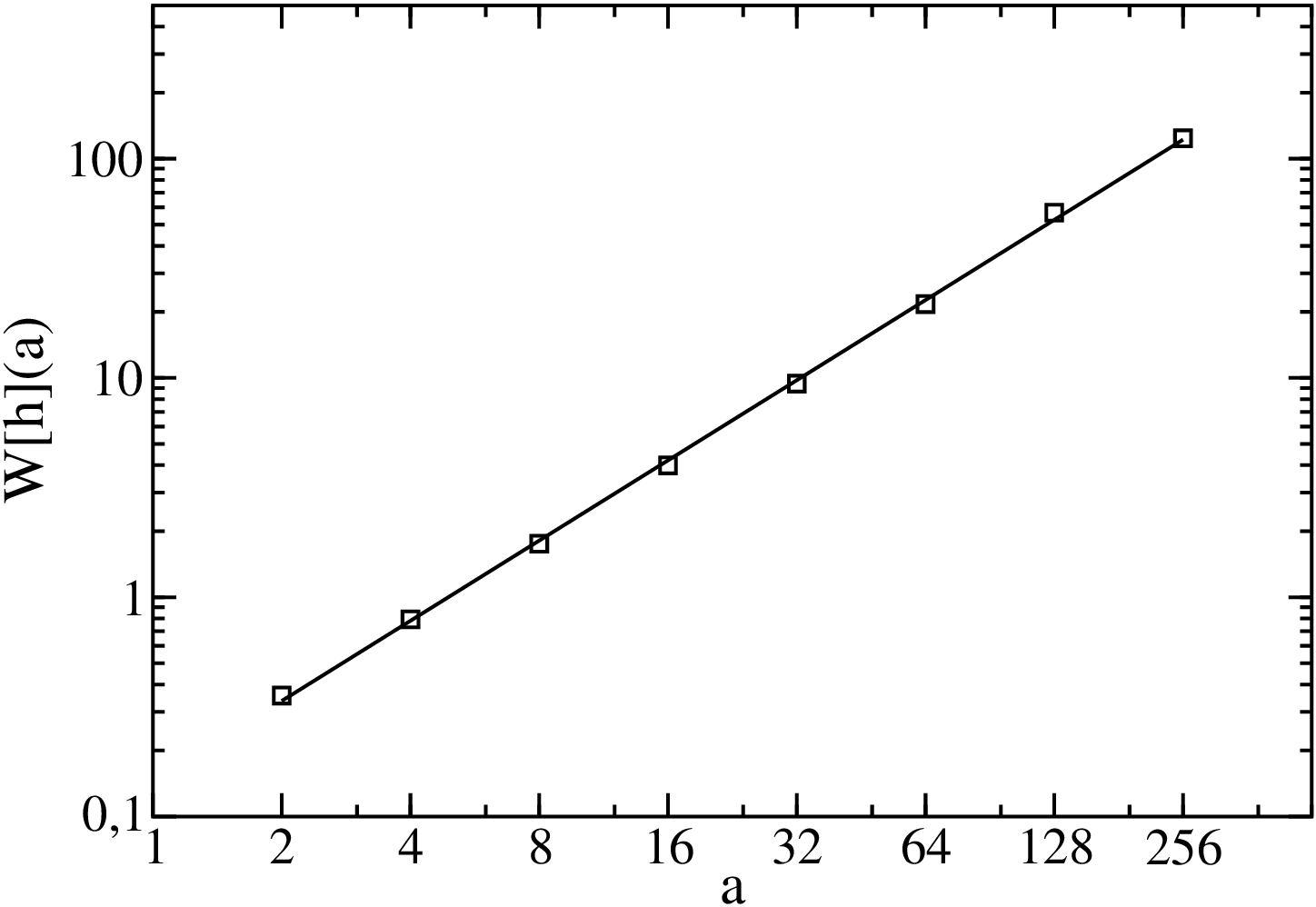}
  \caption{\label{fig:waveletwidth}The local roughness exponent is found 
using the averaged wavelet coefficient (AWC) method. 
The solid line  corresponds to $\zeta_{loc} = 0.72$, see Eq.\ 
(\ref{eq:wascale}).  This analysis was done on 237 
lattices of size $L = 256$ with Daubechies wavelet of order 12. }
\end{figure}

\begin{figure}[tbp!]
\includegraphics[width=7cm]{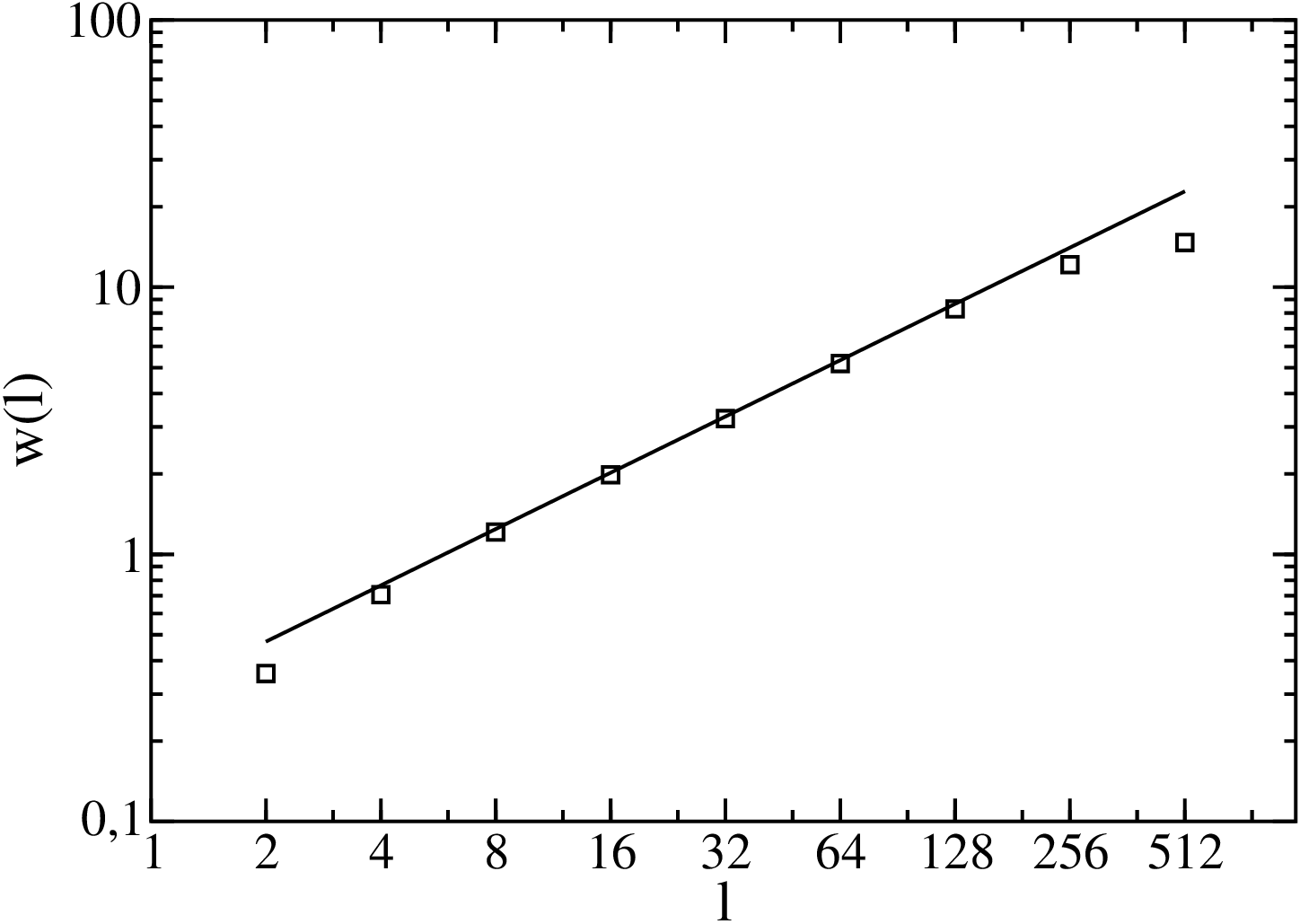}
  \caption{\label{fig:lrmswidth}Local fracture roughness measured by the 
local window method, described in Eq.\ (\ref{eq:wlscale}), 
where the averaged are done over the window 
size $l$ for 237 lattices of system size $L = 256$.  The solid line 
is $\zeta_{loc} = 0.70$.}
\end{figure}

We apply the AWC method and the local window method to fracture profiles of 
lattice size $L = 256$.  We measured a local fracture roughness of 
$\zeta_{loc} = 0.72 \pm 0.02$ for the AWC method and 
$\zeta_{loc} = 0.70 \pm 0.02$ for the local window method, see Figs.\ 
\ref{fig:waveletwidth} and \ref{fig:lrmswidth}.  Both of these methods 
underestimate the roughness exponent, with around 0.03 for the AWC method
and around 0.05 for the local window method.  This was checked by 
creating artificial surfaces with known roughnesses using a wavelet method
for generating the surfaces \cite{sh02}.  
For the local window method the deviations we
found for the measured roughness exponent is consistent with the one found
by Schmittbuhl et al.\ \cite{svr95}. For the AWC method the deviation is
consistent with the one found by Simonsen et al.\ \cite{shn98}
We therefore find the difference between  
$\zeta$ and $\zeta_{loc}$ to be close to zero, which is a sign that anomalous 
scaling, as found by Zapperi et al.\ \cite{zkns05} in the 2D fuse model, 
is not found in the 2D central force model, and that there is only one 
roughness exponent, $\zeta = 0.75 \pm 0.03$, for the central force model.  

Anomalous scaling may be interpreted in the following way.  Assume that the
fracture process has produced a fracture of length $\eta < L$.  The roughness
measured over a window of size $l<\eta$ will be
\begin{equation}
\label{eq:ano}
w(l)=A(\eta)\ l^{\zeta_{loc}}\;,
\end{equation}
where the prefactor $A(\eta)$ depends on $\eta$ as a power law,
\begin{equation}
\label{eq:pref}
A(\eta)\sim \eta^{\zeta-\zeta_{loc}}\;,
\end{equation}
so that $w(l=\eta)\sim \eta^\zeta$.  Imagine we now fix a window size $l < \eta$
and follow the fracture roughness $w(l)$ as $\eta$ grows.  The roughness within
the window changes because the prefactor (\ref{eq:pref}) changes.  This is
not possible unless the cause of the anomalous scaling has its origin in 
statistical non-stationarity \cite{hkm05}, that is, the larger $\eta$ is, the
larger the sample over which the window $l$ is applied to, and this will
change the average on which the prefactor is build.  Using the wavelet-based
method proposed in Ref.\ \cite{hkm05}, we were unable to detect any
such non stationarity, and as a consequence, ruling out anomalous scaling.

\begin{figure}[tbp!]
  \includegraphics[width=7cm]{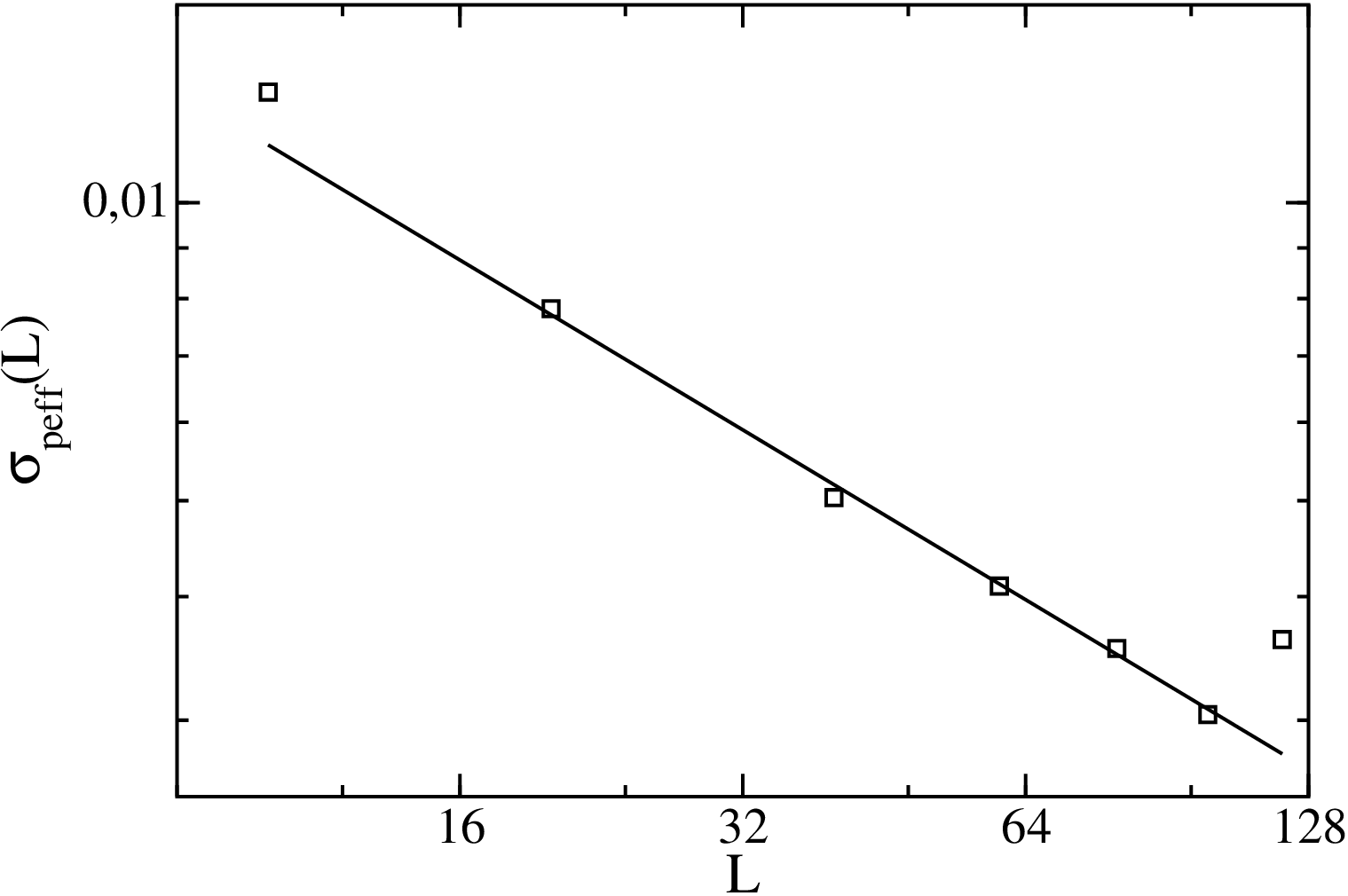}
  \caption{\label{fig:correxp} Log-log plot of the fluctuations in the
density of broken bonds at breakdown.  The solid line is
$1/\nu = 0.57$.  We used from 10000 samples for $L = 10$ to 320 samples
for $L = 100$.}
\end{figure}

When the disorder is broad, the correlation length for the percolation-like 
process $\xi$ is much greater then the system size $L$.  Thus we can 
measure the correlation length exponent by using finite size scaling 
for  the density of broken springs at fracture
\begin{equation}
  p_{eff} = p_c - \frac{C}{L^{1/\nu}}.
\end{equation}
where $p_c$ is the critical density at which the network breaks down in the 
limit $L \rightarrow \infty$, and
$p_{eff}$ is effective critical density at which the network breaks down for 
a finite $L$.
This implies that the fluctuations in the density of broken springs at 
fracture will scale as 
\begin{equation}
  \sigma(p_{eff}) = (\langle p_{eff}^2\rangle - \langle p_{eff}\rangle^2)^{1/2}
 \propto L^{-1/v}\;.
\label{eq:sigma}
\end{equation}
Using a disorder of $D = 20$ which gives flat damage density profiles, 
indicating that there is no localization in the fracture process,
we obtain $1/\nu = 0.57 \pm 0.10$, see Fig.\ \ref{fig:correxp}.  
This corresponds to a value for $\nu$  
equal to $1.7 \pm 0.3$, which by Eq.\ (\ref{eq:sh}) 
gives $\zeta \in \{0.73, 0.80\}$, which is consistent with the direct 
measurement of $\zeta$. The value we found for $\nu$ is different from 
the rigidity percolation value of 1.16 found 
by Moukarzel and Duxbury \cite{md99}. 

To conclude: We have studied the central force breakdown model to test 
Eq.\ (\ref{eq:sh}) and found 
support for it measuring the roughness exponent of 
$\zeta = 0.75 \pm 0.03$. The difference in global and 
local roughness exponent is small and a check for anomalous 
scaling of the wavelet coefficients show no 
sign of such. The correlation length exponent for breakdown process
was found to be $\nu = 1.7 \pm 0.3$ giving consistent 
$\zeta$ values using Eq.\ (\ref{eq:sh}).  The value for $\zeta$ 
and $\nu$ are close to those of the fuse model, suggesting that these 
exponents might be similar for these 
models.    

This work was partially financed by the Norwegian Research Council and the 
computations were done under the auspices of the Norwegian HPC program, NOTUR.


\end{document}